%% file: ms.tex
\documentclass[preprint2]{aastex}
\begin{document}
\title{The fundamental gas depletion and stellar-mass buildup 
  times of star forming galaxies}
\author{Jan Pflamm-Altenburg and Pavel Kroupa}
\affil{Argelander-Institut f\"ur Astronomie (AIfA), Auf dem H\"ugel 71, 
  53121 Bonn, Germany}
\email{jpflamm@astro.uni-bonn.de, pavel@astro.uni-bonn.de}
\begin{abstract}
Stars do not form continuously distributed over star forming
galaxies. They form in star clusters of different masses.
This nature of clustered star formation is taken into account
in the theory of the integrated galactic stellar initial mass function
 (IGIMF) in which the galaxy-wide IMF (the IGIMF)
is calculated by adding all IMFs of young star clusters. 
For massive stars the IGIMF is steeper than the universal
IMF in star clusters and steepens with decreasing SFR which is called
the IGIMF-effect. The 
current SFR and the total H$\alpha$ luminosity of galaxies therefore 
 scale non-linearly
in the IGIMF theory compared to the classical case in which the galaxy-wide 
IMF is assumed to be constant and identical to the IMF in star clusters.
We here apply for the first time the revised SFR-$L_\mathrm{H\alpha}$ relation
on a sample of local volume star forming galaxies with measured H$\alpha$
luminosities. The fundamental results are: i) the SFRs of galaxies scale
linearly with the total galaxy neutral gas mass, ii) the gas depletion time
scales of dwarf irregular and large disk galaxies are 
about 3~Gyr implying that dwarf galaxies do not have lower star formation 
efficiencies than large disk galaxies, and iii) the stellar mass buildup
times of dwarf and large galaxies are only in agreement with downsizing
in the IGIMF context, but contradict downsizing within the  traditional 
framework that assumes a constant galaxy-wide IMF.

\end{abstract}
\section{Introduction}
The determination of accurate current star formation rates (SFRs) is  of
fundamental importance for the understanding and investigation 
of the processes relevant for star formation  and galaxy evolution. A commonly
used tracer to calculate the current SFR of galaxies is the
measurement of the H$\alpha$ emission line which is linked to 
the presence of short-lived massive stars.

The standard way to construct relations between the total H$\alpha$ luminosity
and the total SFR of a galaxy is to apply on galaxy-wide scales
an invariant stellar initial mass function (IMF)
which is determined in star clusters 
\citep[eg.][]{kennicutt1983a,gallagher1984a,kennicutt1994a,kennicutt1998b,kennicutt1998a}.
This method provides a linear SFR-$L_\mathrm{H\alpha}$ relation
 as long as the assumed
galaxy-wide IMF is treated to be constant and independent of the SFR.
 
A basic result of SFR studies is that $SFR/M_\mathrm{gas}$ 
decreases with decreasing total neutral gas mass, $M_\mathrm{gas}$.
Low-gas-mass (dwarf irregular) galaxies turn their gas into stars
over much longer time scales  than  high-gas-mass (large disk) galaxies
\citep[eg.][]{skillman2003a,karachentsev2007a,kaisin2008a}. 
That dwarf galaxies have lower star-formation efficiencies,
defined as $\tau_\mathrm{gas}^{-1}=SFR/M_\mathrm{gas}$,
than massive galaxies has been a generally accepted fact at the base of
most theoretical work on galaxy evolution.

But in the last years it has been shown that the application of the IMF in
star clusters to galaxy-wide scales is doubtful 
\citep{weidner2003a,weidner2005a,weidner2006a,hoversten2008a,meurer2009a,lee2009a}.
Instead, the galaxy-wide IMF has to be calculated by adding all
IMFs of all young star clusters leading to an integrated galactic 
stellar initial mass function (IGIMF). The fundamental property of the
IGIMF-theory is that for massive stars 
it steepens with decreasing total SFR although the shape of the underlying
IMF in each star cluster is universal and constant. This is a direct 
consequence of (i) the star formation process taking place in individual
embedded star clusters on the scale of a pc, 
many of which dissolve when emerging from their molecular
cloud cores,  rather than being uniformly distributed over the galaxy and 
of (ii)  
low-star-formation regions being  dominated by low-mass star clusters which
are void of massive stars \citep*{weidner2003a,weidner2005a,weidner2006a,weidner2009a}.

The main aim of this paper is not to present a high precission data 
analysis of galaxies but to demonstrate the change of fundamental galaxy
properties when switching from the constant IMF assumption 
to the IGIMF theory.

We recap the basics of the 
IGIMF theory and update its observational support 
in Section~\ref{sec_igimf}.
The required data of the sample of star forming galaxies are tabulated in
Section~\ref{sec_data}.  We then calculate the SFRs of galaxies 
(Section~\ref{sec_sfr}), gas depletion time scales (Section~\ref{sec_gas_depl}),
 and stellar mass build-up times (Section~\ref{sec_buildup_times}) 
using an IGIMF based SFR-$L_\mathrm{H\alpha}$ conversion and show
the dramatic differences that arise form using a constant-IMF based
SFR-$L_\mathrm{H\alpha}$ relation.

\section{IGIMF: theory and observation}
\label{sec_igimf}
The IMF within each star cluster seems to be universal 
\citep{elmegreen1997a,elmegreen1999a,kroupa2001a,kroupa2002a,loeckmann2009a}. 
The canonical form
 of the IMF, $\xi(m)=dN/dm$, where $dN$ is the number of stars in the mass
interval $[m,m+dm]$, is a two-part power law,
$\xi(m)\propto m^{-\alpha_\mathrm{i}}$, in the stellar regime
with $\alpha_1 = 1.3$ for $0.1 \le m/\mathrm{M_\odot}<0.5$,
$\alpha_2 = 2.35$ for $0.5 \le m/\mathrm{M_\odot}< m_\mathrm{max}$,
where $m_\mathrm{max}$ is the maximal stellar mass 
in a just-born star cluster with embedded stellar mass $M_\mathrm{ecl}$.
It can also be formulated as a log-normal distribution at low masses
and a power-law extension at high masses \citep{miller1979a,chabrier2003a}.
 
The maximum stellar mass, $m_\mathrm{max}$, up to which the IMF of
a star cluster is populated, is a function of the total stellar mass, 
$M_\mathrm{ecl}$, \citep{weidner2006a}. This relation has been recently
confirmed by an updated much larger sample  of young embedded star clusters
\citep{weidner2009a}. 

The masses of young embedded star clusters are distributed according to the
embedded cluster mass function (ECMF) \citep[eg.][]{lada2003a}. The
ECMF is populated up to the most massive young star cluster, 
$M_\mathrm{ecl,max}(SFR)$. This most massive young star cluster is found 
to scale with the total SFR of the host galaxy \citep*{weidner2004b}.

As stars form in star clusters  the galaxy-wide IMF has to be calculated
by adding all IMFs of all young star clusters. The embedded star cluster
population is dominated by low-mass star clusters and has only a few
high-mass star clusters. But due to the $m_\mathrm{max}$-$M_\mathrm{ecl}$
relation massive stars are predominately formed in high-mass clusters,
whereas low-mass stars are formed in all star clusters.
Thus the resulting integrated galactic stellar initial mass function 
is steeper than the underlying universal canonical IMF in each star cluster.

With decreasing SFR the upper mass limit of the ECMF decreases  and
the number ratio of high-mass to low-mass young star clusters decreases.
Consequently the galaxy-wide number ratio of high-mass to low-mass
young stars decreases, too. As a consequence  the IGIMF steepens with decreasing 
SFR, which we refer to as the IGIMF-effect. 

Among all quantities which are relevant for the IGIMF theory
the slope of the ECMF is the least accurately constrained one. Thus,
two different ECMF-slopes, according to the observational and theoretical 
range of possible slopes, are used to fully explore the possible range
of IGIMF effects: the standard IGIMF with a Salpeter 
single power-law ECMF slope of
$\beta=2.35$, and the minimum-1 IGIMF with a power-law ECMF slope of
$\beta_2=2.00$ for star cluster more massive than 50~$M_\odot$ and 
$\beta_1=1.00$ for star cluster masses between  5 and 50~$M_\odot$.
It has been recently suggested 
that the mass
function of young star clusters might be better described by a 
Schechter-function \citep{bastian2008a,larsen2009a}.
\citet{bastian2008a} obtains  a Schechter function
with a low mass power-law slope of $\beta=2$ below 
$M_\star =$ 1--5 $\times 10^{6}$~M$_\odot$. It follows from the SFR-most massive star cluster relation \citep{weidner2004b} that only galaxies with 
SFR$\gtrsim$30~M$_\odot$~yr$^{-1}$ populate the ECMF beyond the mass regime described by the power law part of the Schechter function. All galaxies in our sample are below this threshould. Thus the power law description of the ECMF in the IGIMF model includes already a possible global Schechter function of the ECMF for these galaxies.

For a deeper introduction the reader is referred to section~2 of 
\citet{pflamm-altenburg2007d}  and references therein.

As a consequence, all galaxy-wide properties which depend predominantly
on the presence of high-mass stars scale non-linearly with the 
total SFR, such as  the total H$\alpha$ luminosity, 
the oxygen yield, and the alpha-element [$\alpha$/Fe] ratios.
The observed mass-metallicty relation of galaxies is a direct
outcome of the IGIMF-theory \citep{koeppen2007a} and the
decreasing [$\alpha$/Fe] ratio with decreasing velocity dispersion
of galaxies can be easily explained in the IGIMF context \citep{recchi2009a}. 
In both cases no extreme fine-tuning and parameter adjustment is required 
as is the case if a constant IMF on galaxy scales is assumed. 

The IGIMF-theory connects the galaxy-wide IMF with the current SFR and thus
  refers to whole galaxies. Within galaxies star formation is described by 
the corresponding surface star formation rate density, 
$\Sigma_\mathrm{SFR} =dM/dt dx dy$, which defines the newly formed stellar 
mass, $dM$, per time interval, $dt$, and per area, $dx dy$. In order to 
construct a local relation between the local star formation rate surface 
density,
$\Sigma_{SFR}$, and the produced local H$\alpha$ luminosity surface density,
$\Sigma_\mathrm{H\alpha}=dL_\mathrm{Halpha}/dt dx dy$, which defines the 
produced H$\alpha$ luminosity, $dL_\mathrm{H\alpha}$, per time interval, $dt$,
and per area, $dx dy$, the embedded cluster mass function and the integrated 
galactic stellar initial mass function are replaced in the IGIMF theory
by their corresponding surface densities. These are
the local embedded cluster mass 
function, LECMF,
and the local integrated galactic stellar initial mass function, LECMF.
In the outer regions of disk galaxies the star formation rate surface density 
is lower and the LECMF is populated to lower upper masses
than in the inner regions. Like the IGIMF effect for whole galaxies, a 
local IGIMF effect emerges. Thus the H$\alpha$ luminosity surface 
density decreases faster with increasing radial distance than the star 
formation rate surface 
denstiy, which naturally explains the radial H$\alpha$ cutoff in disk galaxies
\citep{pflamm-altenburg2008a}.

\begin{figure}
  \plotone{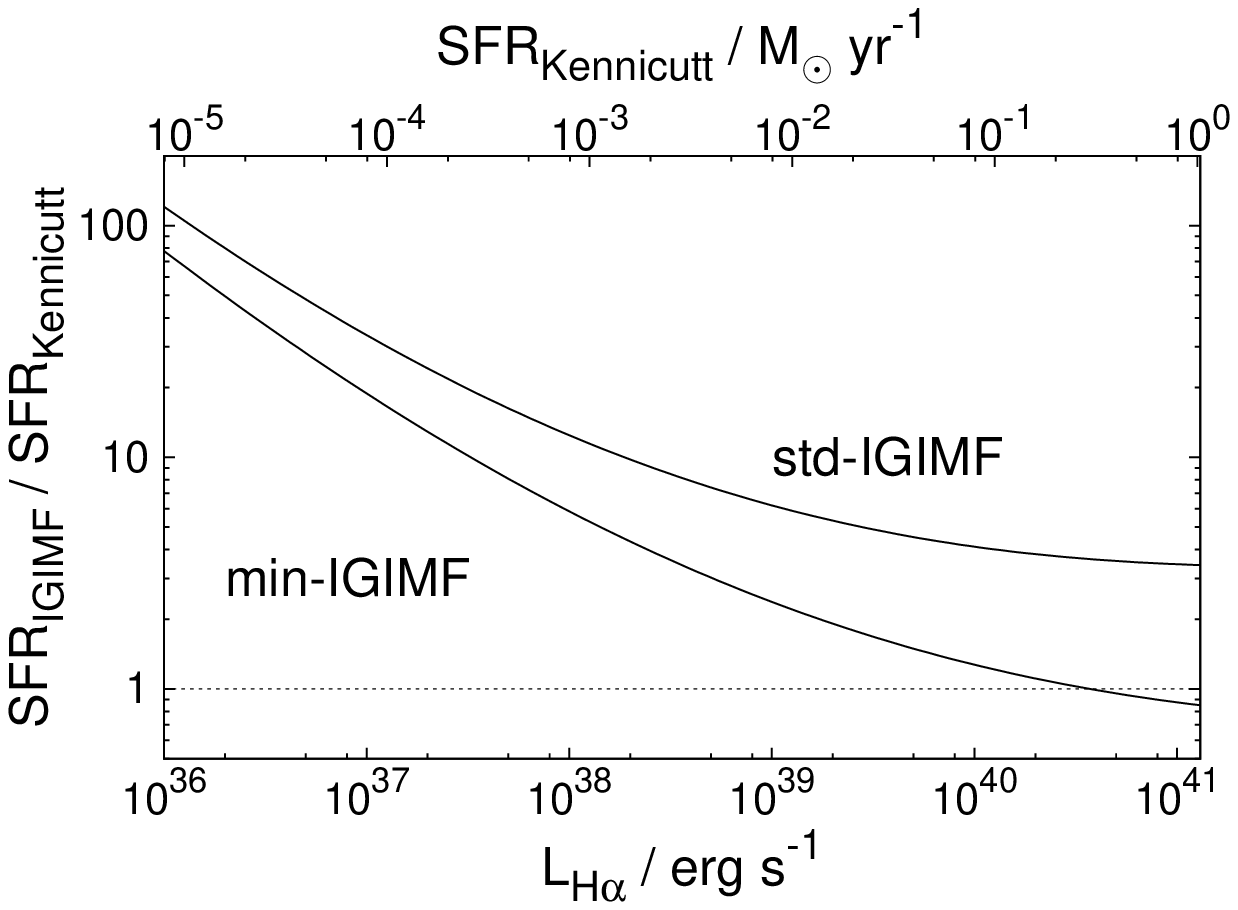}
  \caption{\label{fig_sfr_lha_comp} Ratio of the SFRs calculated from 
    the observed 
    H$\alpha$ luminosity with the classical constant-IMF based conversion from \citet{kennicutt1994a} and the IGIMF based conversion from \citet{pflamm-altenburg2007d}. The IGIMF-based SFRs are calculted using the fith-order polynomal fit
  published in secion~2.3 and tab.~2 in \citet{pflamm-altenburg2007d}.}
\end{figure}

How the H$\alpha$ based SFRs 
of galaxies and their related properties change when
the classical linear $L_\mathrm{H\alpha}$-SFR relation is replaced
by the non-linear IGIMF based conversion, presented in
\citet{pflamm-altenburg2007d}, is explored in this paper.
The classical relation between the total H$\alpha$ luminosity of a galaxy
and the underlying SFR is based on the assumption that the galaxy-wide IMF is identical to the IMF observed in star clusters and does not vary with SFR. Thus, the 
classical SFR scales linearly with the total H$\alpha$ luminosity. The relation between the total SFR and the total H$\alpha$ luminosity in the IGIMF theory is nearly linear in the high H$\alpha$ luminosity range, i.e. for L$_*$  and more luminous galaxies. 
But the IGIMF-based SFR-$L_\mathrm{Halpha}$ relation deviates
increasingly from linearity for H$\alpha$ faint galaxies, i.e. for galaxies with an H$\alpha$ luminosity comparable to the SMC ($\approx 5\times10^{39}$ erg s$^{-1}$) or less. Thus, the classical method underestimates the SFR of H$\alpha$ faint galaxies compared to the IGIMF prediction. This underestimation is
illutrated in Fig.~\ref{fig_sfr_lha_comp}. The SFR for both IGIMF models (standard and minimal) are calculted using the fith-order polynomal fit
  published in secion~2.3 and tab.~2 in \citet{pflamm-altenburg2007d}. The classical SFR is calculated with the widly used linear relation by \citet{kennicutt1994a},
\begin{equation}
  \frac{\mathrm{SFR}}{\mathrm{M_\odot\;yr^{-1}}} = 
  \frac{L_\mathrm{H\alpha}}{1.26\times 10^{41}\;\mathrm{erg\;s^{-1}}}\;.
\end{equation} 
For example, a galaxy with an H$\alpha$ luminosity of $1.26\times 10^{38}$~erg~s$^{-1}$
has a classical Kennicutt-SFR of $10^{-3}$~M$_\odot$~yr$^{-1}$. The IGIMF SFR is 5 times larger for the minimal model and 10 times larger in the standard model.

Additionally the IGIMF-effect is expected
to have different strenghts for H$\alpha$ and FUV radiation, because
the H$\alpha$ luminosity depends only on the presence of the 
ionising radiation of short-lived 
high-mass stars, whereas long-lived B~stars also contribute to 
the FUV flux. It has been predicted that the H$\alpha$/FUV flux ratio
decreases with decreasing SFR \citep{pflamm-altenburg2009a}. 
Recently \citet{meurer2009a} have shown
that the H$\alpha$/FUV flux ratio of star 
forming galaxies decreases with decreasing average H$\alpha$ surface
luminosity density and conclude that a varying galaxy-wide IMF slope
in dependence of the SFR might be the most likely explanation. 
A different study by \citet{lee2009a} of local volume galaxies
found a decreasing H$\alpha$/FUV flux ratio with decreasing total
H$\alpha$ luminosity which is in remarkable quantitative 
agreement with the prediction
by the IGIMF theory by \citet{pflamm-altenburg2009a} down to the
least-massive dwarf galaxies.

A more detailed listing and discussion of the observational 
support of the IGIMF theory is given in section~1 of 
\citet*{pflamm-altenburg2009a}.

It should be emphasised  here that one may argue that anything can 
be explained by a suitably chosen variable IMF. But how the IGIMF varies is 
not adjusted for the problem to be solved or explained. Instead,
the nature of clustered
star formation defines how the slope of the IGIMF scales with the
total SFR. This means the
slope of the IGIMF is defined by the SFR of the galaxy. The relation
between the IGIMF-slope and the current SFR is constant for all applications.
There is no freedom to match the galaxy-wide IMF slope 
for a considered problem. 
Although the slope of the IGIMF varies the IMF in star clusters is universal 
and constant and not varying, i.e. the star formation physics is untouched.
 
\section{Data}
\label{sec_data}
In order to calculate current SFRs, gas depletion times and 
stellar mass buildup time scales of star forming galaxies in the
IGIMF-context, the H$\alpha$ luminosity, $L_\mathrm{H\alpha}$, 
the total neutral gas mass, $M_\mathrm{gas}$, and the blue band absolute magnitude, 
$M_\mathrm{B}$, of 200 local volume galaxies (Tab.~\ref{tab_data})
are compiled from different samples of star forming galaxies: 
the Canes Venatici I
group of galaxies \citep{kaisin2008a}, the M81 galaxy group 
\citep{karachentsev2007a}, the dwarf irregular galaxies of the Sculptor
group \citep{skillman2003a}, a sample of isolated dwarf irregular galaxies
\citep{vanzee2001a}, local group dwarf irregular galaxies \citep{mateo1998a}
and the local group disk galaxies, M31, M33  \citep{walterbos1994a,dame1993a,karachentsev2004a,verley2007a,corbelli1997a,kennicutt1995a,hindman1967,westerlund1997a}. In most cases the H$\alpha$ 
luminosity was not given. Instead, the published SFR, based on the observed 
H$\alpha$ luminosity, has been transformed into the  
H$\alpha$ luminosity using the respective SFR-$L_\mathrm{H\alpha}$ relation
as described in the publication where the SFR is extracted from. 
The published SFR, based on a linear SFR-$L_\mathrm{H\alpha}$ relation
\citep[eg.][]{kennicutt1994a}, 
is tabulated in Table~\ref{tab_data} (Col.~5). The observed total 
H$\alpha$-luminosity (Col.~2) is converted into an IGIMF-SFR
using the relations given in \citet{pflamm-altenburg2007d} for
the standard-IGIMF model (Col.~6) and the minimal1-IGIMF model (Col.~7)
in order to cover the range of IGIMFs allowed by the empirical data 
(Sec.~\ref{sec_igimf}). 

The Milky Way is included in this study. The HI mass is taken from \citet{van_den_bergh1999a}. Due to the lack of H$\alpha$ data for the Milky Way we take the SFR from \citet{diehl2006a} which is based on the meassurement of the Al-26 and Fe-60 gamma ray flux. As this SFR determination is based on a Scalo slope of $\alpha=2.7$ in the high-mass regime of the IMF, the IGIMF-effect is already taken unknowingly into account.  

The total neutral gas mass (Col.~3) is the total HI mass taken from the references 
multiplied by 1.32, following \cite{skillman2003a}, to account for primordial helium. 
The absolute B-band magnitudes are listed in Col.~4.

Other HI--total-gas-mass  conversion factors have been applied in the literature, eg. 1.34 \citep{knuth1986a,legrand2001a} or 1.36 \citep{kennicutt2007a}. 
The  conversion factor may therefore be uncertain by 3~per cent, corresponding to a logarithmic uncertainty of the total neutral gas mass of $\Delta \log_\mathrm{HI} = 0.013$. For the purpose of this study this uncertainty can be 
neglected and we use a factor of 1.32, 
following \citet{skillman2003a}, throughout this study.

Finally it should be noted that the galaxies of our sample stem from 
different observations. Values from 21~cm and optical observations are
compared with each other. As HI exists to much larger galactocentric radii than
H$\alpha$ radiation is produced, the areas covered by the
observations are diffently large. Because our study compares total, integrated 
values with each other the diffently large areas do not impose a problem 
here. The observed H$\alpha$ luminosities might suffer from a bias 
as the different observations might have been performed with differently defined
surface brightness limits. Given  that  the H$\alpha$ surface brightess 
strongly declines with increasing galactocentric radius  
in star forming galaxies, the main contribution to the total 
H$\alpha$ luminosity comes from the inner regions of the galaxies which should
be covered by each observation. The very small possible bias for the total
H$\alpha$ luminosities can be neglected here, because the main aim of our study
is to demonstrate the differences which arise when the classical method for
calculating SFRs from total H$\alpha$ luminosities based on a constant
galaxy-wide IMF is replaced by the IGIMF-based conversion resulting in new
SFRs being up to a factor 100 larger.

In order to detect possible differences between the different data sources 
all galaxies which belong to the same sub-sample
are plotted by the same 
symbol. These sub-samples are the Canes Venatici cloud of galaxies 
(CVnI), the M81 group, the Sculptor group of dwarf irregular galaxies,
a sample of isolated dwarf irregular galaxies,
the local group of star forming galaxies, and the Milky Way.
It can  be seen in the following plots that the changes of the 
SFRs with decreasing galaxy mass are the same for all sub-samples.
A systematic bias between the various sub-samples is thus not evident.

Allthough these sub-samples cover different regions in the local volume,
some overlap exists with the sub-sample of \emph{isolated dwarf irregular 
galaxies} by \citet{vanzee2001a}. Nine galaxies have been identified to be
listed twice in Table~\ref{tab_data} (see Table~\ref{tab_gal_twice}).
These galaxies appear twice in the following plot as independent data.

\include{tab1}
\include{tab2}

\section{Calculating SFRs}
\label{sec_sfr}
\begin{figure}
  \plotone{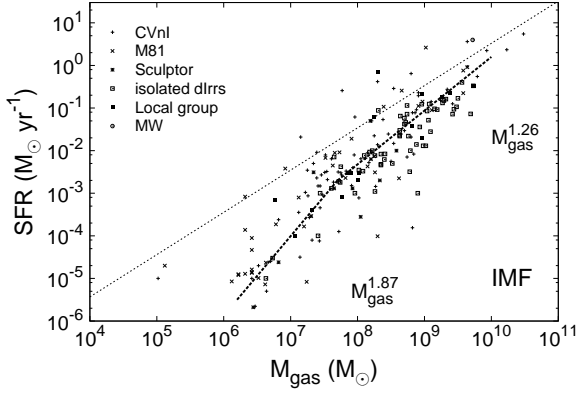}
  \caption{\label{fig_sfr_m_gas_imf}The original SFRs taken from  the 
    literature are plotted versus the host galaxy total neutral gas mass. These
    SFRs are calculated from the total H$\alpha$ luminosity using a classical
    linear SFR-$L_\mathrm{H\alpha}$ relation. Galaxies with a total neutral 
    gas mass 
    $\gtrsim 10^8 M_\odot$ show a SFR$\propto M_\mathrm{gas}^{1.26}$ relation
    (eq.~\ref{eq_sfr_m_gas_imf_high}), 
    whereas the SFRs of less massive galaxies scale 
    with $M_\mathrm{gas}^{1.87}$ (eq.~\ref{eq_sfr_m_gas_imf_low}), 
    both bivariate fits being shown
    as thick-dotted lines.  For comparison, the thin dotted line shows 
    the bivariate regression
    from Fig.~\ref{fig_sfr_m_gas_std-igimf} of the standard IGIMF model
    with slope 0.99.}
\end{figure}

The SFRs, which are  taken from the literature and based on 
a classical linear conversion of the H$\alpha$ luminosity into a SFR
\citep[eg.][]{kennicutt1994a},
are plotted in dependence of the total galaxy neutral gas mass in 
Fig.~\ref{fig_sfr_m_gas_imf}.  With decreasing galaxy neutral gas mass
the relation between the traditionally calculated SFR and the galaxy neutral gas
mass becomes increasingly steeper. Here we find that 
for massive galaxies of our sample with neutral gas masses 
$\ge 5\times10^7\;M_\odot$ the SFR scales slightly
non-linear with neutral gas mass,
\begin{equation}
\label{eq_sfr_m_gas_imf_high}
\frac{SFR}{M_\odot\;\mathrm{yr}^{-1}} = 
3.91\times 10^{-13}\;\left(\frac{M_\mathrm{gas}}{M_\odot}\right)^{1.26}\;,
\end{equation}
and scales 
almost quadratically for less massive galaxies
($M_\mathrm{gas} \le 5\times 10^7\;M_\odot$),
\begin{equation}
  \label{eq_sfr_m_gas_imf_low}
\frac{SFR}{M_\odot\;\mathrm{yr}^{-1}} = 
7.96\times 10^{-18}\;\left(\frac{M_\mathrm{gas}}{M_\odot}\right)^{1.87}\;.
\end{equation}
This behaviour is already well known
\citep[e.g.][]{karachentsev2007a,kaisin2008a}. 

\begin{figure}
  \plotone{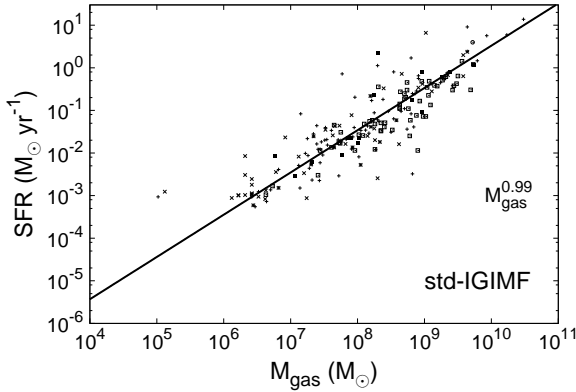}
  \caption{\label{fig_sfr_m_gas_std-igimf}The calculated SFRs based on the 
    standard IGIMF versus the total galaxy neutral gas mass. The solid line shows
    the bivariate regression for the standard IGIMF 
    (eq.~\ref{eq_sfr_m_gas_std_igimf}). The symbols are the same as in Fig.~\ref{fig_sfr_m_gas_imf}.
  }
\end{figure}

\begin{figure}
  \plotone{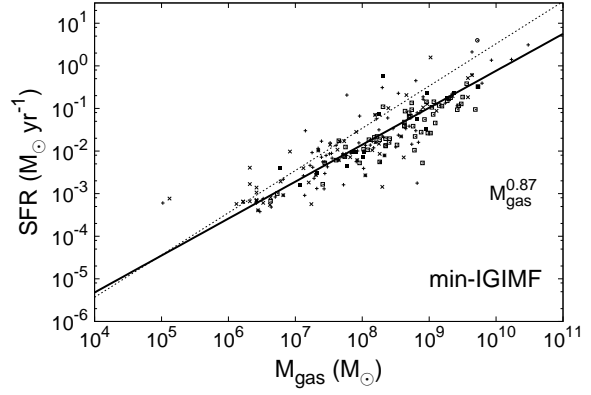}
  \caption{\label{fig_sfr_m_gas_min-igimf}
    Same as Fig.~\ref{fig_sfr_m_gas_std-igimf} but using the 
    SFR-$L_\mathrm{H\alpha}$ relation based on the minimum-IGIMF.
    The solid line shows  the bivariate regression for this minimum IGIMF 
    (eq.~\ref{eq_sfr_m_gas_min_igimf}). For comparison, 
    the dotted line shows the fit of
    the standard model (eq.~\ref{eq_sfr_m_gas_std_igimf}). 
    The symbols are the same as in Fig.~\ref{fig_sfr_m_gas_imf}.}
\end{figure}

The non-linear relation between the total H$\alpha$ luminosity and the
underlying SFR calculated with the IGIMF-theory diverges significantly from
these classical linear relations  based on a constant galaxy-wide IMF below
SFRs typical for SMC-type galaxies. For comparison, the SFR-$M_\mathrm{gas}$
relation in the classical picture becomes increasingly steeper for
galaxy masses less than  SMC-type galaxies. It can be expected
that the IGIMF theory revises the H$\alpha$-SFRs of galaxies with a small
neutral gas content such that their lower star formation efficiencies 
will be increased.
The SFR-$M_\mathrm{gas}$ relation of this galaxy sample as resulting from 
the IGIMF theory is shown in Fig.~\ref{fig_sfr_m_gas_std-igimf}
for the standard model and in Fig.~\ref{fig_sfr_m_gas_min-igimf}
for the minimum model.  

It can be clearly seen that for both  IGIMF models the turn-down 
evident in the classical SFR-$M_\mathrm{gas}$ relation 
(Fig.~\ref{fig_sfr_m_gas_imf}) 
disappears completely. 

For the standard IGIMF a bivariate regression gives a linear scaling
of the total SFR and the galaxy neutral gas mass,
\begin{equation}
\label{eq_sfr_m_gas_std_igimf}
\frac{SFR}{M_\odot\;\mathrm{yr}^{-1}} = 
3.97\times 10^{-10}\;\left(\frac{M_\mathrm{gas}}{M_\odot}\right)^{0.99}\;.
\end{equation} 
For the minimum IGIMF the relation between the SFR and the galaxy 
neutral gas mass is
\begin{equation}
\label{eq_sfr_m_gas_min_igimf}
\frac{SFR}{M_\odot\;\mathrm{yr}^{-1}} = 
1.62\times 10^{-9}\;\left(\frac{M_\mathrm{gas}}{M_\odot}\right)^{0.87}\;.
\end{equation}

Note that the main parameter in the IGIMF theory is the slope of the 
ECMF. Both IGIMF models, the standard model with an ECMF slope of $\beta=2.35$
and the minimum model with an ECMF slope of $\beta=2$, cover the full range
of possible ECMFs.  Therefore, the change that emerges
in the SFR-$M_\mathrm{gas}$ relation when changing from the classical 
invariant IMF description to the IGIMF theory 
is a direct result of the nature  of clustered star formation.

In conclusion, the IGIMF theory revises the SFRs of dwarf galaxies 
significantly, such that the true SFRs may be larger than hitherto thought
by two or three orders of magnitude. The reason why this was not
evident until now comes about because in dwarf galaxies star formation 
predominantly occurs in low-mass clusters which do not contain massive
stars.


\section{Gas depletion time scales}
\label{sec_gas_depl}
\begin{figure}
  \plotone{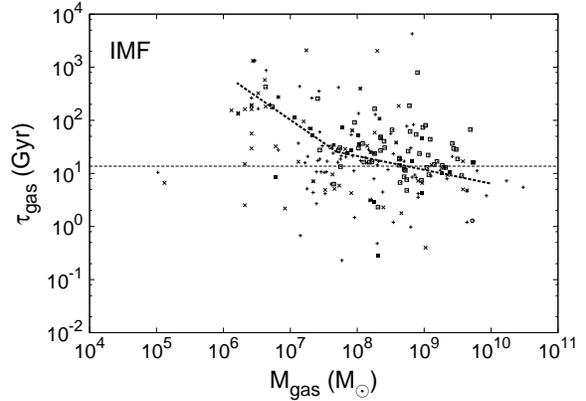}
  \caption{\label{fig_t_gas_depl__m_gas_imf} Gas depletion time scales 
    in dependence of the total galaxy neutral gas mass for the case of
    a constant galaxy-wide IMF. The thin dotted line marks the age 
    of the universe,  13.7~Gyr. The symbols are the same as in Fig.~\ref{fig_sfr_m_gas_imf}.
  }
\end{figure}

 The gas depletion time scale, $\tau_\mathrm{gas}$, is a measure
 for how long 
a galaxy with current total neutral gas mass, $M_\mathrm{gas}$, 
can sustain its current SFR without being refueled by accretion 
of fresh extra-galactic material. It is defined by
\begin{equation}
  \tau_\mathrm{gas} = \frac{M_\mathrm{gas}}{SFR}\;.
\end{equation}
The corresponding gas depletion time scales, based on a constant 
galaxy-wide IMF, are plotted in Fig.~\ref{fig_t_gas_depl__m_gas_imf}.

The SFR-$M_\mathrm{gas}$ relations for the IMF case 
(eq.~\ref{eq_sfr_m_gas_imf_high} and \ref{eq_sfr_m_gas_imf_low}) can 
now be converted into  $\tau_\mathrm{gas}$-SFR relations, which are plotted
as dashed lines in Fig.~\ref{fig_t_gas_depl__m_gas_imf}.
The gas depletion time scale increases slightly with decreasing total
galaxy neutral gas mass for galaxies more massive than $5\times 10^7$~$M_\odot$ 
and it increases strongly for less-massive galaxies with masses less than 
$5\times 10^7$~$M_\odot$
, in agreement with other studies 
\citep*{skillman2003a,bothwell2009a}. This is commonly interpreted
as star forming dwarf galaxies having much lower star formation 
efficiencies, $\tau_\mathrm{gas}^{-1}$, than large disk galaxies. 

In the previous section it has been shown that the SFRs of dwarf galaxies
are much higher if the constant galaxy-wide IMF is replaced by the 
IGIMF, and consequently the gas depletion time scales are now expected 
to be shorter. 

\begin{figure}
  \plotone{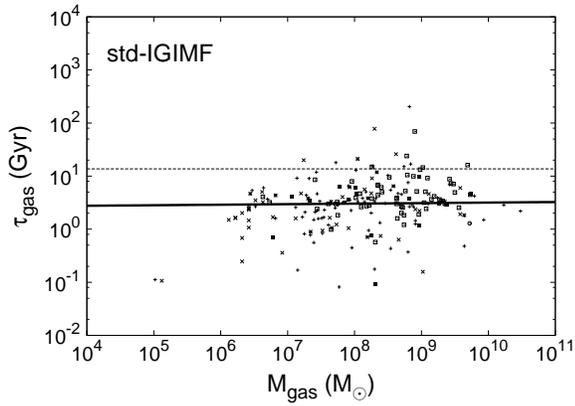}
  \caption{\label{fig_t_gas_depl__m_gas_std-igimf} Same as 
    Fig.~\ref{fig_t_gas_depl__m_gas_imf} but for the
    case of the standard IGIMF. The solid line shows 
    eq.~\ref{eq_tau_gas_std_igimf}, while the thin 
    dotted line marks the age of the universe as in 
    Fig.~\ref{fig_t_gas_depl__m_gas_imf}. Note the reduced scatter
    in comparison to Fig.~\ref{fig_t_gas_depl__m_gas_imf}.
    The symbols are the same as in Fig.~\ref{fig_sfr_m_gas_imf}.
   }
\end{figure}

\begin{figure}
  \plotone{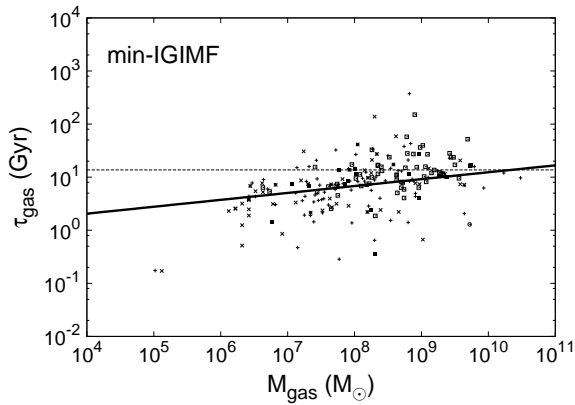}
  \caption{\label{fig_t_gas_depl__m_gas_min-igimf}
    Same as 
    Fig.~\ref{fig_t_gas_depl__m_gas_imf} but for the
    case of the minimum IGIMF. The solid line shows 
    eq.~\ref{eq_tau_gas_min_igimf}.
    The symbols are the same as in Fig.~\ref{fig_sfr_m_gas_imf}.
   }
\end{figure}

The resulting gas depletion times scales for the 
standard IGIMF are plotted in Fig.~\ref{fig_t_gas_depl__m_gas_std-igimf}
and for the minimum IGIMF in Fig.~\ref{fig_t_gas_depl__m_gas_min-igimf}.
In both cases the gas depletion time scales of dwarf galaxies are comparable
to the gas depletion time scales of large disk galaxies. 

For the standard IGIMF the SFR scales linearly 
with the total galaxy neutral gas mass (eq.~\ref{eq_sfr_m_gas_std_igimf}). The resulting
gas depletion times scale with the total galaxy neutral gas mass as
\begin{equation}
  \label{eq_tau_gas_std_igimf}
  \tau_\mathrm{gas} = \frac{M_\mathrm{gas}}{SFR} = 
  2.52\;\mathrm{Gyr}\,\left(\frac{M_\mathrm{gas}}{M_\odot}\right)^{0.01}\;.
\end{equation}
The mean gas depletion time scale of a galaxy with a 
neutral gas mass of $10^6~M_\odot$  is 2.89~Gyr and for a galaxy with a neutral gas
mass of $10^{10}~M_\odot$ it is 3.17~Gyr. Thus, 
the gas depletion time scales are constant for all galaxies, being 
approximately 3~Gyr. It follows that  low-mass dwarf irregular 
galaxies have the same star formation efficiency as massive 
disk galaxies, in disagreement with the currently widely accepted notion
according to which dwarf galaxies are inefficient in making stars.

\begin{figure}
  \plotone{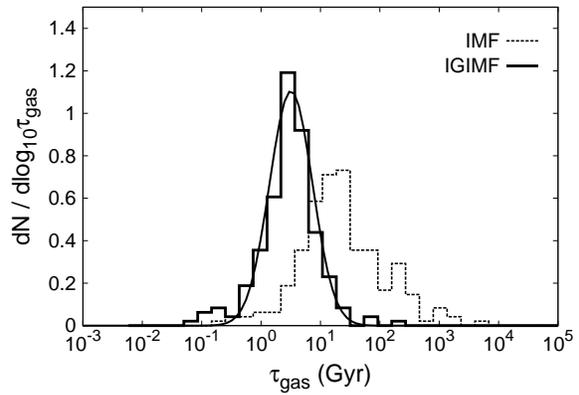}
  \caption{\label{fig_t_gas_depl_histo_std-igimf} The normalised
    histogram of gas depletion
    time scales for the constant IMF (Fig.~\ref{fig_t_gas_depl__m_gas_imf}) 
    and the standard IGIMF (Fig.~\ref{fig_t_gas_depl__m_gas_std-igimf}) 
    overlayed with the log-normal distribution of eq.~\ref{eq_log_norm}.
    Note that the IGIMF-based gas depletion time scales are significantly
    shorter on average, have a symmetric distribution,
    and have a significantly smaller dispersion than if
    the galaxy-wide IMF is assumed to be invariant.}
\end{figure}

The gas depletion time scales for the standard IGIMF are not only 
shorter than the IMF gas depletion time scales, but they also have
a smaller scatter. Fig.~\ref{fig_t_gas_depl_histo_std-igimf} 
shows the normalised histogram of the gas depletion time scales
for the IMF case (dashed histogram) and for the standard IGIMF case
(solid histogram). The scatter of the standard IGIMF gas depletion time
scales can  be well described by a log-normal distribution,
\begin{equation}
\label{eq_log_norm}
  \frac{dN_\mathrm{Gal}}{d log \tau_\mathrm{gas}}=
  \frac{1}{\sqrt{2\pi\sigma^2}}\;
  e^{-\frac{\left(\log\tau_\mathrm{gas}-\mu\right)^2}{2\sigma^2}}\;,
\end{equation}
with $\mu=9.50$ (3.19~Gyr), $\sigma=0.36$. If and to what degree
conclusions can be made for the physical conditions for star formation 
on the basis of the reduced scatter in the IGIMF gas depletion time scales
is unclear at the moment. A volume limited sample in further studies
 is required in order to avoid any kind of bias.

 In the case of the minimum IGIMF the resultant 
$\tau_\mathrm{gas}$-$M_\mathrm{gas}$ relation follows from 
eq.~\ref{eq_sfr_m_gas_min_igimf},
\begin{equation}
  \label{eq_tau_gas_min_igimf}
  \tau_\mathrm{gas} = \frac{M_\mathrm{gas}}{SFR} = 
0.62\;\mathrm{Gyr}\,\left(\frac{M_\mathrm{gas}}{M_\odot}\right)^{0.13}\;.
\end{equation} 
This leads to a gas-depletion time scale of 3.74~Gyr for a $10^6$~$M_\odot$ 
galaxy and 12.37~Gyr for a $10^{10}$~$M_\odot$ galaxy. This means that in 
the case of the minimum IGIMF dwarf irregular galaxies consume their
gas faster than large disk galaxies and would therefore have
higher star-formation efficiencies in contradiction to the 
currently widely accepted picture.

Thus, independent of the IGIMF model details dwarf galaxies 
do not have lower star
formation efficiencies than large disk galaxies.

\section{Blue stellar mass buildup times}
\label{sec_buildup_times}
In the previous section the current SFR and available total neutral gas mass
of galaxies have been analysed to calculate their current
gas depletion time scales. This allows an estimation of their future
star formation activity. Here we now compare the current SFR, and the
already assembled total stellar mass, $M_*$, of galaxies to analyse their past 
star formation by calculating the corresponding stellar mass buildup
time scale, $\tau_*$, which is defined by
\begin{equation}
\label{eq_tau_star}
\tau_* = \frac{M_*}{SFR}\;.
\end{equation}

The total stellar mass is here calculated from the blue-band magnitude,
$M_\mathrm{B}$, (Tab.~\ref{tab_data}, Col.~4). In order to account
for possible  different metallicities and star formation histories (SFHs),  
we use the stellar-mass-to-light-ratio--colour relations published 
by \citet{bell2001a}.
These relations are obtained from synthetic spectra based on  different
galaxy evolution models with a  constant galaxy-wide IMF.  \citet{bell2001a}
adopt a large variety of star formation histories as indicated in 
Tab.~\ref{tab_M_L}.

\include{tab3}

It might be argued that the application of these relations is wrong because the 
IGIMF will affect the stellar-mass-to-light-ratio--colour relations.  
In \citet{pflamm-altenburg2009a} it has been shown that for small SFRs
the IGIMF effect
for the FUV-flux is less than one dex if the IGIMF effect for
H$\alpha$ is two dex. 
The B-band and longer wavelength bands are expected to show
an even smaller IGIMF effect than the FUV-flux. 
The dominating influence on the stellar
mass buildup time scale is by far the IGIMF-effect for the H$\alpha$
based SFRs (eq.~\ref{eq_tau_star}). 
Thus, the \citet{bell2001a} stellar-mass-to-light-ratio--colour
relations are applicable for an order-of-magnitude estimate of the 
stellar mass buildup times. However, precision galaxy evolution modeling would
require revised mass-to-light ratios, which we will adress in future projects.

The two stellar-mass-to-light-ratio--colour relations 
\begin{equation}
  \log\frac{M_\mathrm{*}}{L_\mathrm{B}} = a_\mathrm{B,B-V} 
  + b_\mathrm{B,B-V}\left(M_\mathrm{B}-M_\mathrm{V}\right)\;, 
\end{equation}
and
\begin{equation}
  \log\frac{M_\mathrm{*}}{L_\mathrm{B}} = a_\mathrm{B,V-H} 
  + b_\mathrm{B,V-H}\left(M_\mathrm{V}-M_\mathrm{H}\right)\;, 
\end{equation}
tabulated in \citet{bell2001a} can be used to construct
a stellar-mass-to-light-ratio--colour relation between the B-band magnitude
and the \hbox{B-H}-colour,
\begin{equation}
  \label{eq_B_B-H}
  \log\frac{M_\mathrm{*}}{L_\mathrm{B}} = a_\mathrm{B,B-H} 
  + b_\mathrm{B,B-H}\left(M_\mathrm{B}-M_\mathrm{H}\right)\;,
\end{equation}
where the two coefficients are given by
\begin{equation}
  \label{eq_aa_b}
  a_\mathrm{B,B-H}=
  \frac{b_\mathrm{B,V-H}\;a_\mathrm{B,B-V}+b_\mathrm{B,B-V}\;a_\mathrm{B,V-H}}
       {b_\mathrm{B,V-H}+b_\mathrm{B,B-V}}\;,
\end{equation}
and
\begin{equation}
  \label{eq_a_b}
  b_\mathrm{B,B-H}=
  \frac{b_\mathrm{B,V-H}\;b_\mathrm{B,B-V}}{b_\mathrm{B,V-H}+b_\mathrm{B,B-V}}\;,
\end{equation}

The required colour is obtained using the tabulated blue-band magnitude,
and the H-band magnitude, $M_\mathrm{H}$. To assign an H-band magnitude
to each galaxy we use the observed extremely tight correlation between the 
B- and H-band magnitude of galaxies \citep[eq.~9]{kirby2008a} ranging from 
$M_\mathrm{B}=-8$ to $M_\mathrm{B}=-22$. The resulting empirical 
relation between
the B-H colour and the B-band is then given by
\begin{equation}
  \label{eq_B_H}
  M_\mathrm{B}-M_\mathrm{H} = -0.14\;M_\mathrm{B}+0.74\;.
\end{equation}

In \citet{bell2001a} the stellar-mass-to-light-ratio--colour relations
are given for seven different galaxy evolution models, which lead to slightly
different values for $a_\mathrm{B,B-V}$, $a_\mathrm{B,V-H}$, 
$b_\mathrm{B,B-V}$, and $b_\mathrm{B,V-H}$. The resulting
coefficients, $a_\mathrm{B,B-H}$ and $b_\mathrm{B,B-H}$, 
are listed in Tab.~\ref{tab_M_L}. In the following analysis we use their
mean values for the 
$\log M_\mathrm{*}/L_\mathrm{B}$--($M_\mathrm{B}-M_\mathrm{H}$) relation,
\begin{equation}
  \label{eq_a_b_values}
  a_\mathrm{B,B-H} = -1.633\;\;\;\mathrm{and}\;\;\; b_\mathrm{B,B-H}=0.602\;.
\end{equation}

We now have for each of our galaxies $M_\mathrm{B}$ (Tab.~\ref{tab_data}, Col~4),
$M_\mathrm{H}$ (from eq.~\ref{eq_B_H}) and thus $M_\mathrm{*}/L_\mathrm{B}$
(from eq.~\ref{eq_B_B-H}).
Each of the $M_\mathrm{B}$ values can therewith be converted to the total
stellar mass $M_\mathrm{*}$ allowing the computation of $\tau_\mathrm{*}$
for each galaxy.

For a few galaxies of our sample absolute H-band magnitudes are listed
in \citet{kirby2008a}. For these galaxies we use the observed absolute H-band
magnitude, for all the others we use the calculated B-H colour according to the
procedure described above. Table~\ref{tab_H_obs_cal} 
lists those galaxies which have  observed
absolute H-band magnitudes by \citet{kirby2008a}.
\include{tab4}

We now estimate the error of the calculated stellar mass 
due to different galaxy evolution models and theoretical stellar-mass-to-light 
ratios. Consider one galaxy with observed blue-band luminosity, $L_\mathrm{B}$, 
and current star formation rate, SFR. Two different theoretical 
stellar-mass-to-light ratios, $\Upsilon_1$ and $\Upsilon_2$, lead to two 
different stellar masses, $M_{*,1}$ and $M_{*,2}$. The logarithmic difference of the 
two corresponding stellar mass
buildup times, $\tau_{*,1}$ and $\tau_{*,2}$, is
\begin{eqnarray}\nonumber&&
  \log\tau_{*,2}-\log\tau_{*1} = \log\frac{\tau_{*,2}}{\tau_{*,1}}=
  \log\frac{M_{*,2}\; SFR}{SFR\;M_{*,1}}\\
  \nonumber&&
  =\log\frac{M_{*,2}\; SFR\;L_\mathrm{B}}{L_\mathrm{B}\;SFR\;M_{*,1}}=
  \log\frac{\Upsilon_{2}}{\Upsilon_1}\\
  &&
  =\log\Upsilon_{2}-\log\Upsilon_1\;.
\end{eqnarray}
Note that $L_\mathrm{B}$ and $SFR$ can be crossed out, because they refer
to the same galaxy.

\begin{figure}
  \plotone{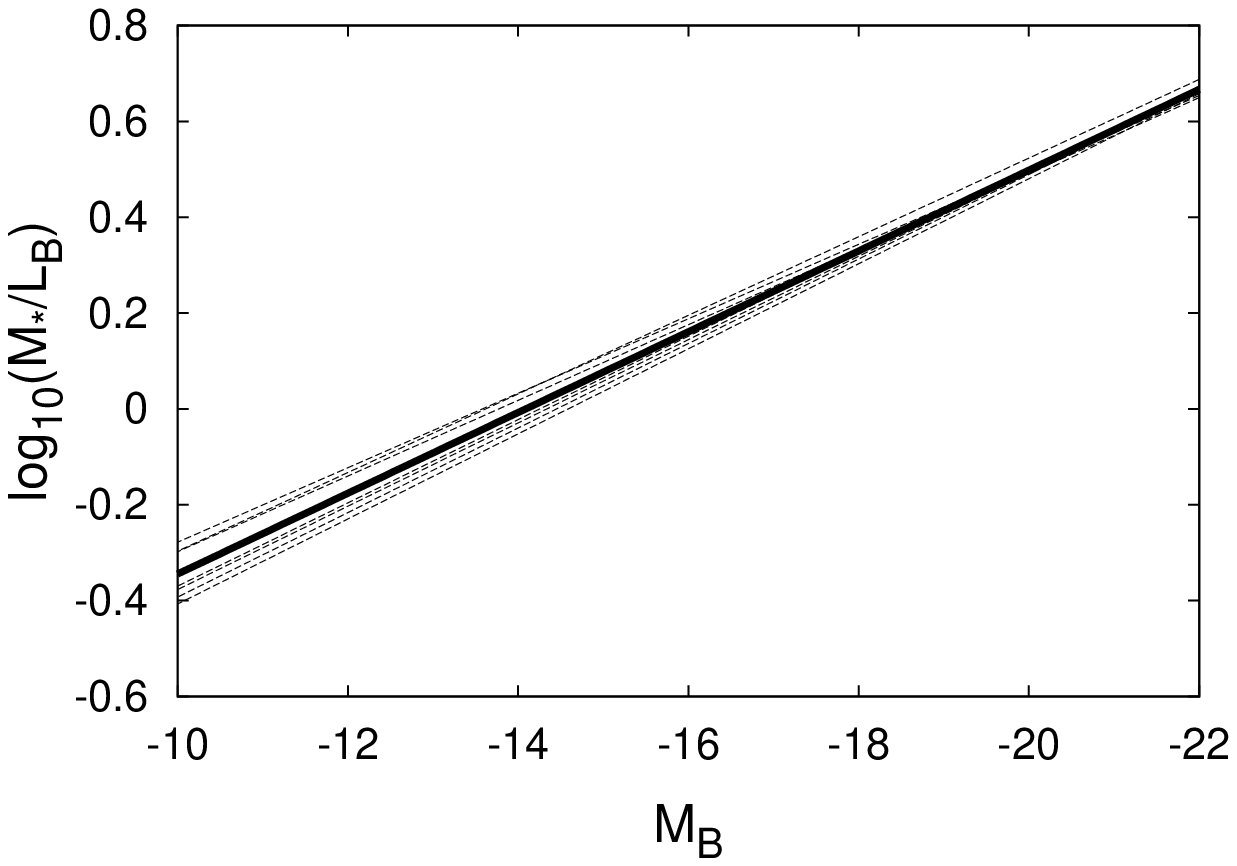}
  \caption{\label{fig_M_L_error} The stellar-mass-to-light ratio as a 
    function of the blue-band magnitude resulting from combining
    eq.~\ref{eq_B_H} and \ref{eq_B_B-H} with Tab.~\ref{tab_M_L}.
    Each dotted line represents one individual galaxy evolution model
    and the solid line shows the mean value relation. Note that the increase
    of $M_\mathrm{*}/L_\mathrm{B}$ with $M_\mathrm{B}$ is not a result 
    of the IGIMF effect but occurs as a result of using the empirical 
    relation (eq.~\ref{eq_B_H}) in the constant-IMF modelling of 
    \citet{bell2001a}.
  }
\end{figure}

In Fig.~\ref{fig_M_L_error} the stellar-mass-to-light ratio is plotted
as a function of the total blue-band magnitude for the different galaxy
evolution models listed in Tab.~\ref{tab_M_L}. For the least massive
galaxies the model variation in the stellar-mass-to light ratio is
about $\pm 0.05$ dex around the mean value relation and becomes smaller for
more massive galaxies. Therefore, for the purpose of this analysis 
the uncertainty in the calculation of the current stellar mass due 
to galaxy evolution can be neglected.

\begin{figure}
  \plotone{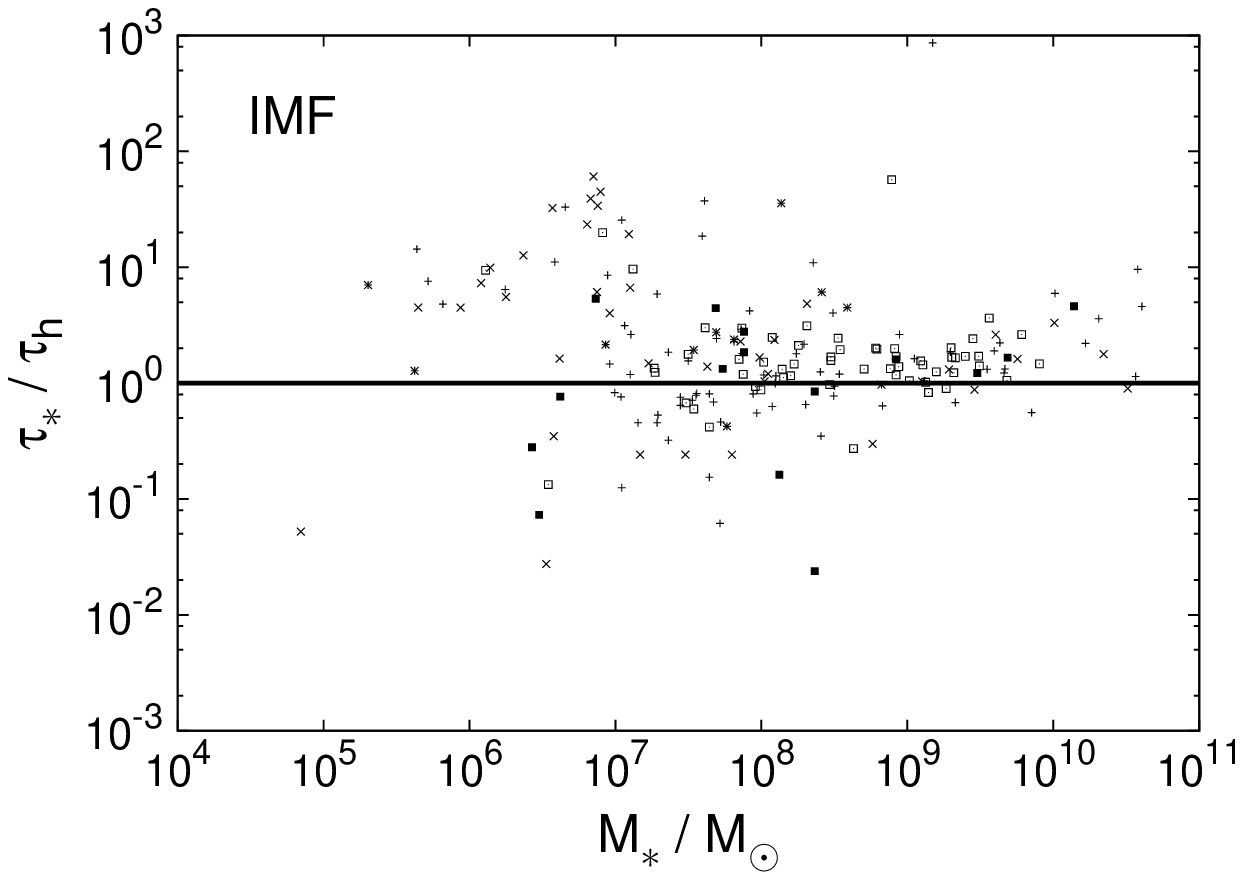}
  \caption{\label{fig_t_star_m_star_imf}
    The stellar mass buildup time scale in units of the Hubble time, 
    $\tau_\mathrm{h}$~=~13.7~Gyr, in dependence of the total stellar mass
    derived from the blue luminosity for a constant galaxy-wide IMF. Note
    that the buildup times for dwarf star forming galaxies are longer than
    the ones for large disk galaxies. This implies that the SFRs of dwarf 
    galaxies must have decreased faster over cosmic time than the SFRs
    of large disk galaxies. This is in contradiction to the finding of
    downsizing according to which the SFRs of large disk galaxies must have decreased
    faster over cosmic time than the SFRs of dwarf galaxies.
    The symbols are the same as in Fig.~\ref{fig_sfr_m_gas_imf}.
  }
\end{figure}

For the case of a constant galaxy-wide IMF the resulting stellar-mass
buildup times as a function of the total stellar mass are
plotted in Fig.~\ref{fig_t_star_m_star_imf}. The average value of the
stellar-mass buildup time is constant for galaxies more massive than
$10^8~M_\odot$ and increases by more than one order of magnitude for
less massive ones. This implies that the SFR in massive galaxies has
decreased much more slowly over cosmic time than for the least 
massive galaxies. This
is in contradiction to the finding of downsizing, according to which 
massive disk galaxies have on average older stellar populations than 
dwarf galaxies.

\begin{figure}
  \plotone{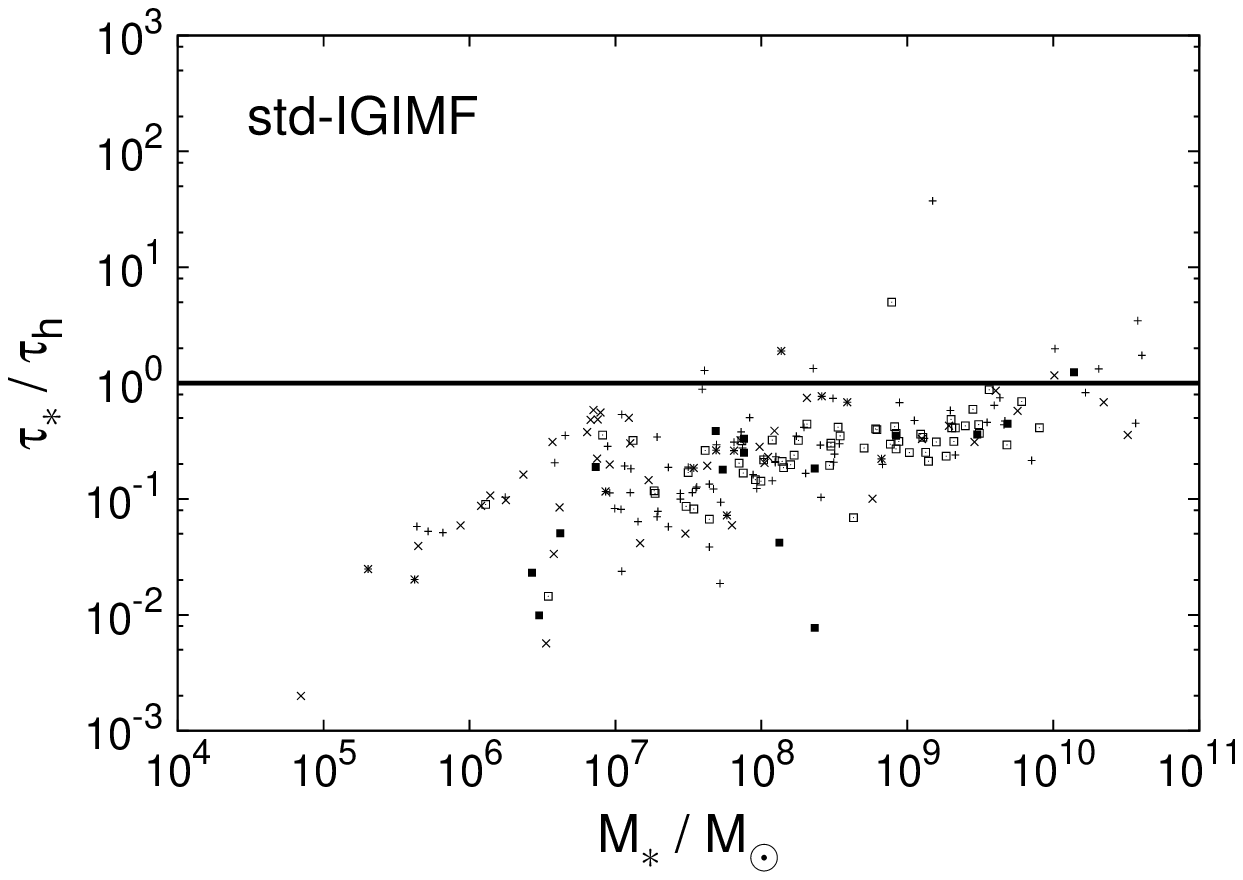}
  \caption{\label{fig_t_star_m_star_std_igimf}
    The same as Fig.~\ref{fig_t_star_m_star_imf} but for the standard IGIMF.
    Note that the stellar mass build-up times, which result from SFR 
    calculations in the IGIMF context, decrease with decreasing galaxy mass.
    This suggest that the SFR may have been increasing slighly with time 
    for dwarf galaxies or that they were forming stars over a more recent epoch
    while for massive disk galaxies the SFHs may be constant, 
    which is in agreement with downsizing.
    The symbols are the same as in Fig.~\ref{fig_sfr_m_gas_imf}.
  }
\end{figure}

\begin{figure}
  \plotone{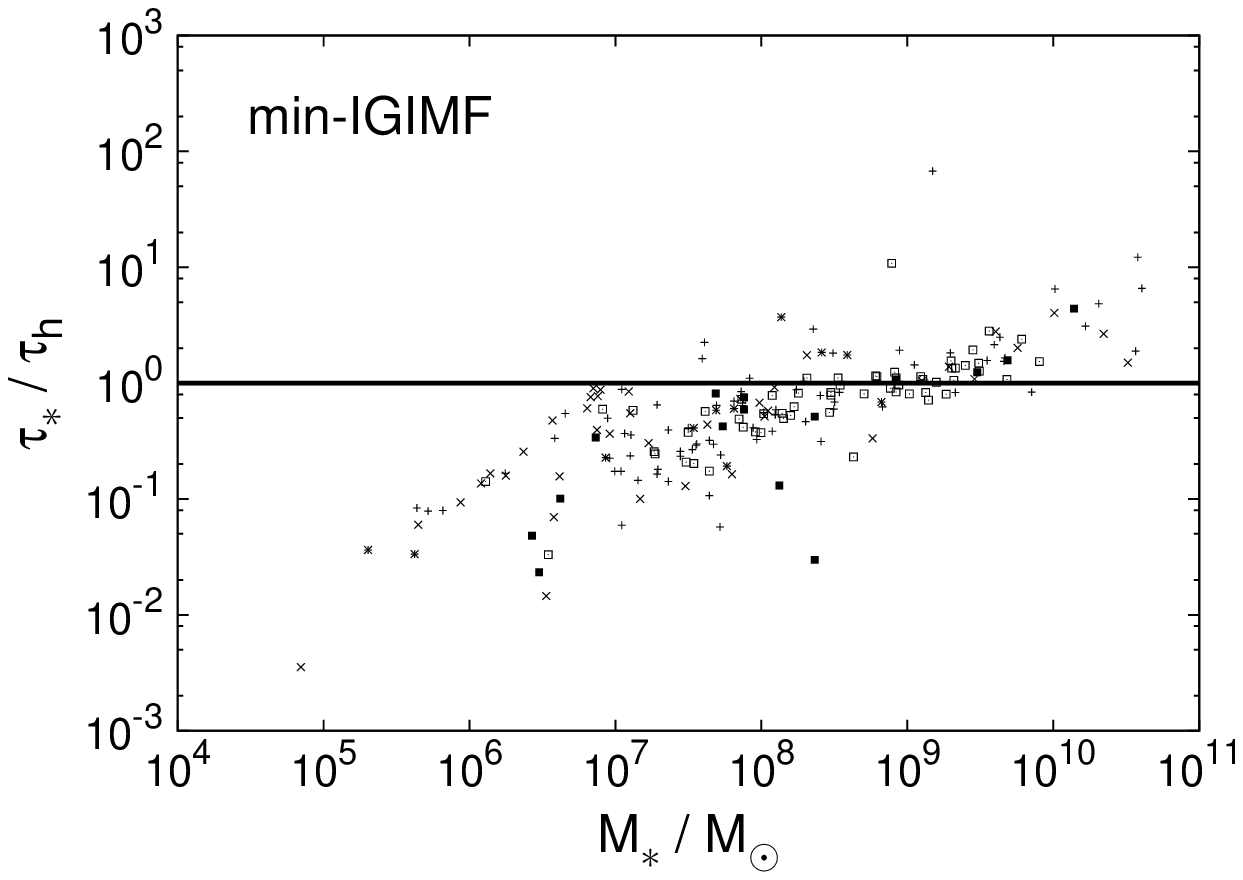}
  \caption{\label{fig_t_star_m_star_min_igimf}
    The same as Fig.~\ref{fig_t_star_m_star_std_igimf} but for the minimum IGIMF.
    The symbols are the same as in Fig.~\ref{fig_sfr_m_gas_imf}.
  }
\end{figure}

When using H$\alpha$ as a SFR tracer,
in the IGIMF context the SFRs are much higher than the constant IMF-based
SFRs for low SFR galaxies, i.e. less massive galaxies. The corresponding
stellar-mass buildup times therefore become shorter. Indeed, for both 
IGIMF models, standard (Fig.~\ref{fig_t_star_m_star_std_igimf}) 
and minimum (Fig.~\ref{fig_t_star_m_star_min_igimf}), 
the stellar-mass buildup times
decrease monotonically with decreasing galaxy stellar mass.
This suggests that massive
disk galaxies have been forming stars at about a constant rate for a Hubble
time, while dwarf galaxies have either turned on more recently or have
a slightly increasing SFH, in full agreement with downsizing.

\section{Conclusion}

We have applied the revised SFR-$L_\mathrm{H\alpha}$
relation, which takes the nature of clustered star formation into account,
to a significantly larger sample of star forming galaxies of the local 
volume than was available in \citet{pflamm-altenburg2007d}.

The SFRs of galaxies and the corresponding gas depletion and
stellar-mass buildup time scales are calculated. 
Comparing these new values with the
classical ones which are derived from SFRs based on an assumed constant 
galaxy-wide IMF, the changes are dramatic:

The SFRs of galaxies scale linearly with the total galaxy neutral gas mass and the
corresponding  gas depletion time scales are independent of the galaxy
neutral gas mass. This implies that dwarf galaxies have the same star formation 
efficiencies as large disk galaxies. Furthermore, the stellar-mass
buildup times are only compatible with downsizing in the IGIMF context.
They are inconsistent with downsizing in the classical theory which
assumes an invariant galaxy-wide IMF. 

These results follow from the 
SFR$_\mathrm{H\alpha}$/SFR$_\mathrm{FUV}$ data compiled by \citet{lee2009a},
independently of the well developed IGIMF theory. The IGIMF theory merely
explains the \citet{lee2009a} results within an empirically well established
star-formation framework. The success of the IGIMF theory,
developed by \citet{weidner2003a} and \citet{weidner2005a,weidner2006a}, is
that the \citet{lee2009a} results have been predicted 
\citep{pflamm-altenburg2007d} rather than being
adjusted afterwards. The power of the IGIMF theory lies in that it
is readily computable and deterministic.

We emphasise that the galaxy properties derived in the IGIMF theory 
are qualitatively 
independent of the main parameter of the IGIMF, the slope of the ECMF. 
For the full range of 
possible slopes of the ECMF in the IGIMF theory 
SFRs of dwarf galaxies are significantly higher than derived in the classical
paradigm assuming a constant galaxy-wide IMF independent
of the global SFR.

The revision of the SFRs of dwarf galaxies suggested here 
now poses a major challenge for our theoretical 
understanding of star formation in galaxies which had been developed with the
aim of explaining the low star formation efficiencies of dwarf galaxies.
It follows that galaxy and cosmic evolution models need a substantial revision 
requiring further studies.  

\bibliographystyle{aa}
\bibliography{imf,OB-star,star-formation,cmf,stellar-evolution,stellar-spectra,sfh,disk_galaxies,galaxy-evolution,star-cluster,astro-ph}

\end{document}

%% file: tab1.tex
\begin{deluxetable}{cccccccccc}
  \tabletypesize{\footnotesize}
  \tablewidth{0pt}
  \tablecaption{\label{tab_data}Galaxy properties}
  \label{tab_sculptor-dIrr}
\tablehead{
  \colhead{Galaxy}&
  \colhead{$\log L_\mathrm{H\alpha}$}&
  \colhead{$\log M_\mathrm{gas}$}&
  \colhead{$M_\mathrm{B}$}&
  \colhead{$\log \mathrm{SFR}$}& 
  \colhead{$\log \mathrm{SFR}$}&
  \colhead{$\log \mathrm{SFR}$}&
  \colhead{Ref.}&
  \colhead{Ref.}&
  \colhead{Ref.}\\
  &  erg / s & M$_\odot$ & & M$_\odot$ / yr & M$_\odot$ / yr & M$_\odot$ / yr 
  & SFR, $L_\mathrm{H\alpha}$ & 
  $M_\mathrm{HI}$ & $M_{B}$\\
  &  & & & IMF & std-IGIMF & min-IGIMF & & & }
\startdata
UGC~5427   & 38.78 & 7.61 & -14.48 & -2.19 & -1.47 & -1.87 & (1) & (1) & (1)\\
UGC~5672   & 38.87 & 7.52 & -14.65 & -2.10 & -1.40 & -1.81 & (1) & (1) & (1)\\
NGC~3274   & 39.86 & 8.86 & -16.16 & -1.11 & -0.60 & -1.10 & (1) & (1) & (1)\\
NGC~3344   & 40.71 & 9.79 & -19.03 & -0.26 & 0.17 & -0.41 & (1) & (1) & (1)\\
UGC~6541   & 39.29 & 7.15 & -13.71 & -1.68 & -1.08 & -1.52 & (1) & (1) & (1)\\
NGC~3738   & 39.67 & 8.21 & -16.61 & -1.30 & -0.77 & -1.25 & (1) & (1) & (1)\\
NGC~3741   & 38.69 & 8.17 & -13.13 & -2.28 & -1.53 & -1.92 & (1) & (1) & (1)\\

\enddata
\tablecomments{Table~\ref{tab_data} is published in its entirety in the
electronic edition of the Astrophysical Journal. A portion is shown here
for guidance regarding its form and content.
References:  (1) \citet{kaisin2008a}, (2) \citet{karachentsev2007a},
(3) \citet{skillman2003a}, (4) \citet{mateo1998a}, (5) \citet{vanzee2001a},
(6) \citet{walterbos1994a}, (7) \citet{dame1993a}, 
(8) \citet{karachentsev2004a},
(9) \citet{verley2007a}, (10) \citet{corbelli1997a}, 
(11) \citet{kennicutt1995a}, 
(12) \citet{hindman1967}, (13) \citet{westerlund1997a}, 
(14) \citet{diehl2006a}, (15) \citet{van_den_bergh1999a}.}
\end{deluxetable}

%% file: tab2.tex
\begin{deluxetable}{ccccccc}
  \tabletypesize{\footnotesize}
  \tablewidth{0pt}
  \tablecaption{\label{tab_gal_twice}Multiply listed galaxies}
\tablehead{
  \colhead{Galaxy (ref.)}&
  \colhead{Galaxy (ref.)}&\\
}
\startdata
UGCA 292 (1) & UGCA  292 (5)\\
DDO 181  (1) & UGC 8651 (5)\\
DDO 183  (1) & UGC 8760 (5)\\
UGC 8833 (1) & UGC 8833 (5)\\
DDO 187  (1) & UGC 9128 (5)\\
DDO 190  (1) & UGC 9240 (5)\\
UGC 5423 (2) & UGC 5423 (5)\\
DDO 226  (3) & UGCA 9   (5)\\
DDO 6    (3) & UGCA 15  (5)\\
\enddata
\tablecomments{Each line refers to one galaxy which is listed in different 
  references. The references are the 
  same as in Table~\ref{tab_data}. For example, the second line shows that the
  same galaxy is listed as DDO~181 in reference (1) and as UGC~8651 in 
  reference (5).}
\end{deluxetable}

%% file: tab3.tex
\begin{deluxetable}{lcc}
  \tabletypesize{\normalsize}
  \tablewidth{0pt}
  \tablecaption{\label{tab_M_L}Stellar-mass-to-light-ratio--colour relations}
\tablehead{
  \colhead{Model}&
  $a_\mathrm{B,B-H}$&
  $b_\mathrm{B,B-H}$
}
\startdata
Closed Box                 &-1.551    &0.586\\
Infall                     &-1.706    &0.621\\
Outflow                    &-1.466    &0.555\\
Dynamical time             &-1.505    &0.564\\
Formation epoch            &-1.741    &0.630\\
Formation epoch burst      &-1.764    &0.634\\
\citealt{cole2000a}          &-1.699    &0.621\\
\hline
Mean                       &-1.633    &0.602\\
\enddata
\tablecomments{The coefficiants $a_\mathrm{B,B-H}$ (eq.~\ref{eq_aa_b})
and $b_\mathrm{B,B-H}$ 
(eq.~\ref{eq_a_b}) have been derived from different galaxy evolution models 
\citep{bell2001a} and the corresponding mean values.}
\end{deluxetable}

%% file: tab4.tex
\begin{deluxetable}{ccccccc}
  \tabletypesize{\footnotesize}
  \tablewidth{0pt}
  \tablecaption{\label{tab_H_obs_cal}Observed vs. calculated H-band magnitudes}
\tablehead{
  \colhead{Galaxy}&
  \colhead{$M_\mathrm{H,obs.}$}&
  \colhead{$M_\mathrm{B,obs.}-M_\mathrm{H,obs.}$}&
  \colhead{$M_\mathrm{B,obs.}-M_\mathrm{H,cal.}$}&
  \colhead{$\log M_{*,obs.}$}& 
  \colhead{$\log M_{*,cal.}$}&
  \colhead{$M_{*,obs.}/M_{*,cal.}$}\\
  \colhead{}&
  \colhead{}&
  \colhead{}&
  \colhead{}&
  \colhead{$M_\odot$}&
  \colhead{$M_\odot$}&
  \colhead{}\\
}
\startdata
SC 18 & -11.5 & 2.30 & 2.03  &   5.62 &  5.46 &  1.45795\\
ESO 473-G24 & -16.1 & 3.42 & 2.52  &   7.69 &  7.15 &  3.50501\\
SC 24 & -10.7 & 2.31 & 1.91  &   5.31 &  5.07 &  1.72994\\
NGC 625 & -19.2 & 2.89 & 3.02  &   8.82 &  8.90 &  0.831176\\
ESO 245-G05 & -17.2 & 1.61 & 2.92  &   7.76 &  8.55 &  0.162112\\
DDO 210 & -12.9 & 1.94 & 2.27  &   6.11 &  6.31 &  0.629059\\
ESO 347-G17 & -18.3 & 3.51 & 2.81  &   8.59 &  8.17 &  2.63657\\
ESO 348-G09 & -17.2 & 3.45 & 2.67  &   8.14 &  7.66 &  2.96873\\

\enddata
\tablecomments{{Listed} are those galaxies of Table~\ref{tab_data} for which
absolute H-band magnitudes, $M_\mathrm{H,obs}$, are published in 
\citet{kirby2008a}. Two B-H colors and two total stellar masses based on 
eq~\ref{eq_B_B-H} and \ref{eq_a_b_values} are calculated: $M_\mathrm{B,obs.}-M_\mathrm{H,obs.}$ and $\log M_{*,obs.}$  are based on the observed absolute B-band
(Table~\ref{tab_data}) and the observed absolute H-band magintude 
  \citep{kirby2008a}. $M_\mathrm{B,obs.}-M_\mathrm{H,cal.}$ and $\log M_{*,cal.}$
are based on the observed absolute B-band
(Table~\ref{tab_data}) and the calculated B-H color using eq.~\ref{eq_B_H}.}
\end{deluxetable}